# Axon diameters and myelin content modulate microscopic fractional anisotropy at short diffusion times in fixed rat spinal cord

Noam Shemesh

Champalimaud Neuroscience Programme, Champalimaud Centre for the Unknown, Lisbon, Portugal

*Corresponding author:

Dr. Noam Shemesh, Champalimaud Neuroscience Programme, Champalimaud Centre for the Unknown, Av. Brasilia 1400-038, Lisbon, Portugal

E-mail: noam.shemesh@neuro.fchampalimaud.org;

Phone number: +351 210 480 000 ext. #4467

Running title: Microstructural correlates of microscopic anisotropy

Figures: 8

Tables: 3

# Abstract

Mapping tissue microstructure accurately and noninvasively is one of the frontiers of biomedical imaging. Diffusion Magnetic Resonance Imaging (MRI) is at the forefront of such efforts, as it is capable of reporting on microscopic structures orders of magnitude smaller than the voxel size by probing restricted diffusion. Double Diffusion Encoding (DDE) and Double Oscillating Diffusion Encoding (DODE) in particular, are highly promising for their ability to report on microscopic fractional anisotropy (µFA), a measure of the pore anisotropy in its own eigenframe, irrespective of orientation distribution. However, the underlying correlates of µFA have insofar not been studied. Here, we extract µFA from DDE and DODE measurements at ultrahigh magnetic field of 16.4T in the aim to probe fixed rat spinal cord microstructure. We further endeavor to correlate µFA with Myelin Water Fraction (MWF) derived from multiexponential $T_2$ relaxometry, as well as with literature-based spatially varying axonal diameters. In addition, a simple new method is presented for extracting unbiased µFA from three measurements at different b-values. Our findings reveal strong anticorrelations between µFA (derived from DODE) and axon diameter in the distinct spinal cord tracts; a moderate correlation was also observed between µFA derived from DODE and MWF. These findings suggest that axonal membranes strongly modulate µFA, which – owing to its robustness towards orientation dispersion effects – reflects axon diameter much better than its typical FA counterpart. The µFA exhibited modulations when measured via oscillating or blocked gradients, suggesting selective probing of different parallel path lengths and providing insight into how those modulate µFA metrics. Our findings thus shed light into the underlying microstructural correlates of µFA and are promising for future interpretations of this metric in health and disease.

# Introduction

Diffusion Magnetic Resonance Imaging (MRI) has become a mainstay of contemporary microstructural imaging in biomedical applications. Diffusion MRI can provide rich information on the sample's microstructure by interrogating micron-scale dimensions within millimeter-scale voxels (Johansen-Berg, H, Behrens, 2009). In the hierarchical scaling of dimensions in biological systems, the micron-scale is fortuitously a characteristic length scale of many (sub)cellular structures of interest, such as axons, dendrites or cell bodies, which cannot be accessed using routine spatial resolutions in MRI. Most diffusion MRI methods utilize variants of Stejskal and Tanner's (Stejskal, E.O., Tanner, 1965) Single Diffusion Encoding (SDE) technique (Shemesh et al., 2016), which probes diffusion using a single diffusion epoch spanned by diffusion-sensitizing gradient waveforms. The flexibility of SDE in terms of parameter space led to numerous variants (Grebenkov, 2007), as well as diffusion models (Assaf et al., 2013; Panagiotaki et al., 2012), that have been devised to probe different aspects of the microstructure. For example, Diffusion Tensor Imaging (DTI) models diffusion using a single tensor (Basser and Jones, 2002; Mori and Zhang, 2006) under the assumption of (time-dependent) Gaussian diffusion, and the tensor's rotationally invariant properties can then report on diffusion anisotropy and parallel/perpendicular diffusivities. Other methods, such as q-space imaging (Callaghan et al., 1991; Cohen and Assaf, 2002) or diffusion spectrum imaging (Wedeen et al., 2005) utilize Fourier relationships between the diffusion propagator and signal decay with the q-value (where $\boldsymbol{q} = \frac{1}{2\pi}\gamma\delta\boldsymbol{G}$ is the wavevector, $\gamma$ is the gyromagnetic ratio, $\delta$ represents the gradient duration, and $|\mathbf{G}|$ is the gradient amplitude) to extract information on pore size or orientation distributions, respectively. Diffusion time- and/or frequency-dependence can also provide much insight into the restricting geometry by probing the way in which the diffusion path is modulated with time and/or the diffusion spectrum, respectively (Clark et al., 2001; Fieremans et al., 2016; Gore et al., 2010; Jespersen et al., 2017a; Latour et al., 1994; Novikov et al., 2011; Stepišnik et al., 2006; Veraart et al., 2016a). Furthermore, more advanced biophysical modeling has been recently put forth to characterize specific microstructural components such as neurite density (Jespersen et al., 2007, 2010), or water fractions tentatively associated with axons in white matter (Veraart et al., 2016a) from specific acquisition schemes. Such SDE methods have been widely useful in neuroscience (Zatorre et al., 2012) and biomedical applications, typically targeting longitudinal processes such as stroke, learning, or chronic disease progression (Johansen-Berg, H, Behrens, 2009).

One interesting metric that can be probed by diffusion is the microscopic diffusion anisotropy (µA) (Callaghan and Komlosh, 2002; Cheng and Cory, 1999; Mitra, 1995), from which its normalized counterpart – the microscopic fractional anisotropy (µFA) – can be derived. µFA defines a single compartment's anisotropy in its own eigenframe (Jespersen et al., 2013), e.g., for a sphere µFA = 0 while for an infinite cylinder µFA can approach 1. However, in practice, the MRI signal will always originate from an ensemble, thereby making it necessary to account for orientation dispersion within the ensemble (Jespersen et al., 2017c). In systems comprising coherently-aligned anisotropic objects where orientation dispersion is ideally zero, µFA would be equivalent to the fractional anisotropy (FA) derived from DTI. However, in conventional SDE methods, when orientation dispersion is significant, estimated FA values typically do not represent the true anisotropy, or µFA, as they are conflated with orientation dispersion (Mollink et al., 2017; Reisert et al., 2017). For example, in ideal randomly oriented infinite cylinders, the averaging of anisotropic compartments results in FA = 0, which – without a-priori knowledge or extensive modelling – would suggest that the microscopic geometry is spherical.

In recent years, the Double Diffusion Encoding (DDE) methodology (Figure 1) has been gaining increasing attention for its potential to refine and identify microstructural aspects not so easily probed by SDE (Cory, DG, Garroway, AN, Miller, 1990; Mitra, 1995). Unlike SDE, DDE probes diffusion correlations using – as its name suggests – two diffusion encoding periods, spanned by two independent gradient wavevectors, which are separated by a mixing time ($\tau_m$). Comparing q-space-like signal decays using parallel and perpendicular relative gradient orientations, Cheng and Cory have been able to measure the sizes of randomly oriented elongated (anisotropic) yeast cells, and distinguish them from spherical cells (Cheng and Cory, 1999). Similarly, Callaghan and Komlosh have shown that diffusivities extracted from parallel vs. perpendicular DDE could provide insight into µFA in randomly oriented liquid crystals characterized by Gaussian diffusion (Callaghan and Komlosh, 2002). Such measurements provided the first clues that µFA (termed using many divergent terms (Shemesh et al., 2016)) could be recovered from DDE irrespective of orientation dispersion.

Mitra (Mitra, 1995), and later Özarlsan (Özarslan, 2009) derived exact solutions for DDE signals, and have identified the importance of the mixing time in decoupling µA from other effects. In the short mixing time regime, interesting diffusion-diffraction phenomena can be produced (Laun et al., 2011, 2012, Shemesh et al., 2010a, 2010b, 2012b), and angular dependencies can provide insight into pore sizes as shown experimentally first by Koch and Finsterbusch (Koch and Finsterbusch, 2008, 2011) and then by others (Komlosh et al., 2011; Morozov et al., 2015; Shemesh et al., 2009); however, by analyzing the displacement correlation tensor (Nørhøj Jespersen and Buhl, 2011), the short $\tau_m$ angular DDE experiment was found by Jespersen to be equivalent to a time-dependent SDE experiment (Jespersen, 2012). By contrast, in the long mixing time regime, µA is decoupled from these restriction effects, making its measurement much less complicated (Mitra, 1995; Özarslan, 2009). The ability to measure accurate µA values was validated in (Shemesh et al., 2010a) and its importance was shown in biological systems such as ex-vivo neural tissues (Shemesh and Cohen, 2011a), yeast cells (Shemesh et al., 2011), and preclinical in-vivo experiments (Shemesh et al., 2012a), where the orientational variance of the measurements was highlighted. Lawrenz et al have proposed rotationally invariant schemes for mapping an index of µA (Lawrenz et al., 2010; Lawrenz and Finsterbusch, 2015), and Jespersen et al subsequently generalized rotationally invariant DDE measurements up to $5^{th}$ order (in q-values) via a measurement scheme termed DDE 5-design (Jespersen et al., 2013). Numerous promising studies have also been performed on human scanners (Avram et al., 2013; Finsterbusch, 2011; Koch and Finsterbusch, 2008, 2011; Lawrenz and Finsterbusch, 2015; Ulloa et al., 2015), suggesting quite promising potential for disentangling µFA from the underlying orientation dispersion. Additional recent experiments have even extended the DDE methodology towards MR spectroscopy, aiming to impart specificity towards specific cell populations via cellular-specific metabolites (Shemesh et al., 2014a, 2017).

As alluded to above, the diffusion process in biological tissues is highly time-dependent, and thus the filter with which the diffusion experiment is performed can be important. Oscillating Diffusion Encoding (ODE) experiments (Does et al., 2003; Gore et al., 2010; Stepišnik, 1993) have been widely used in SDE to enhance contrast in neural tissue, likely since they access shorter diffusion time than could be reached using pulsed-gradient-spin-echo methods (Drobnjak et al., 2016).

Additionally, ODE has been shown to be highly beneficial for mapping axonal sizes in rat spinal cord (Xu et al., 2009, 2014) as well as for contrasting malignancy in tissues (Reynaud et al., 2016; Xu et al., 2012). More recently, the DDE framework was extended towards accommodation of oscillating gradients, termed Double Oscillating Diffusion Encoding (DODE, Figure 1A), first in theory (Ianuş et al., 2017b), and more recently, in experiment (Ianuş et al., 2017a). Importantly, DODE enables the time/frequency-dependence of µFA to be studied. Furthermore, DODE sequences reach the long mixing time regimes much more easily than their DDE counterparts, thereby making the experiments less mixing-time dependent (Ianuş et al., 2017b), and, as a result, offering the benefit of reduced echo times. This property is likely due to the mixing beginning already from the first gradient pair, and accumulating over the entire gradient waveform. Such DODE experiments were recently reported for the first time in the ex-vivo mouse brain, and µFA maps derived from DODE indeed showed richer contrast than those of their DDE-derived counterparts (Ianuş et al., 2017a).

Many studies have investigated the underlying microstructural correlates of FA, mainly in white matter (for a classical review, the reader is referred to (Beaulieu, 2002)). It is clear that although myelin strongly modulates FA, it is not necessary for detection of anisotropy in biological systems. Axonal membranes, for example, can impede the diffusion processes with orientational preference and thus can contribute to FA. However, in most studies attempting to investigate the origins of restriction in tissues, orientation dispersion was conflated with SDE-driven metrics; an interesting question is therefore whether µFA, which should not suffer from orientation dispersion effects, could be associated with microstructural features to different extents than FA. *The goal of this study was therefore to investigate how µFA and FA correlate underlying microstructural features such as myelin water fraction (MWF) or axonal diameters.* As well, we aimed to investigate whether these parameters are differently correlated, to qualitatively assess the importance of orientation dispersion, especially in the white matter. The final goal of this study was to determine whether µFA is modulated when different length scales are probed via DODE and DDE sequences. A well-characterized system, namely, fixed spinal cord – which has been extensively used in the past to study diffusion (Jespersen et al., 2017b; Klawiter et al., 2011; Komlosh et al., 2008; Schwartz et al., 2005; Xu et al., 2014) or relaxation (Kozlowski et al., 2008a, 2008b; Nunes et al., 2017; Wilhelm et al., 2012) microstructural correlates – was used for these investigations. Our findings demonstrate interesting differences in correlations between µFA and FA and MWF, as well as with the a-priori known axonal sizes in white matter, when measured using DODE or DDE. Interesting findings in gray matter tissues are also presented. Implications for D(O)DE contrasts and future routes for investigations of the origin of µFA in neural tissue, are discussed.

## Theory

Most MRI studies up to date have used only a single b-value to extract µFA from DDE experiments. However, very recently, Ianus et al showed that for most plausible microstructural scenarios, µFA obtained in such a way can be highly biased due to neglecting the higher-order terms in the signal decay (Ianuş et al., 2017a). Ianus et al proposed to more accurately estimate µFA in both DDE and DODE methodologies by performing D(O)DE experiments at multiple b-values, and fitting both µA (from which µFA is then calculated) and the higher-order term via polynomial fits. That is, the D(O)DE signal decay at long mixing times can be expanded with b-value as:

[Eq. 1]

$$\frac{1}{12}\sum \log(S_\parallel(b)) - \frac{1}{60}\sum \log(S_\perp(b)) = \mu A^2 b^2 + P_3 b^3,$$

where $\mu A^2 = \frac{3}{5}\text{var}(\sigma_i)$, $\sigma_{i=1,2,3}$ are the diffusion tensor eigenvalues, $S_\parallel$ and $S_\perp$ represent the D(O)DE signals acquired using parallel and perpendicular gradients, respectively, and $P_3$ contains the higher-order terms. Ianus et al showed that polynomial fitting can be used to estimate $\mu A^2$ and $P_3$ from Eq. 1. When the mean diffusivity (MD) is additionally measured at lower b-values (e.g., from fitting a tensor to the 12 parallel orientations in the 5-design), µFA can be directly calculated from Eq. 2:

[Eq. 2]

$$\mu FA = \sqrt{\frac{3}{2}\frac{\mu A^2}{\mu A^2 + \frac{3}{5}MD^2}}.$$

Although polynomial fitting probably yields the more accurate estimates of µA², it should be noted that ideally, many b-value shells would be required for robust fitting. An alternative approach would be to acquire a much more minimalistic dataset and still be able to quantify µA² and $P_3$. Setting $\frac{1}{12}\sum \log(S_\parallel(b)) - \frac{1}{60}\sum \log(S_\perp(b)) \equiv \tilde{\epsilon}(b)$, Eq. 1 can be rewritten for two different b-values $b_1$ and $b_2$:

[Eq. 3]

$$\begin{cases}\tilde{\epsilon}(b_1) = \mu A^2 b_1^2 + P_3 b_1^3 \\ \tilde{\epsilon}(b_2) = \mu A^2 b_2^2 + P_3 b_2^3\end{cases}.$$

It is then straightforward to show that from two measurements at different b-values, µA² can be directly obtained from

[Eq. 4]

$$\widetilde{\mu A^2} = \frac{\tilde{\epsilon}(b_2) - \tilde{\epsilon}(b_1)\frac{b_2^3}{b_1^3}}{b_2^2 - \frac{b_2^3}{b_1}},$$

which can then be plugged into Eq. 2 to obtain µFA directly. Note that we use the tilde to distinguish the extracted $\widetilde{\mu A^2}$ from the real $\mu A^2$. This approach for accurate µFA extraction thus requires, in principle, only two measurements, one at low b-value, from which MD and $\tilde{\epsilon}(b_1)$ would be obtained, and another at higher b-value, where $\tilde{\epsilon}(b_2)$ would be obtained. However, since at low b-values required for accurate estimation of MD, $\tilde{\epsilon}(b_1)$ may be very small and comparable to noise levels, it is more appropriate to acquire $\tilde{\epsilon}(b_1)$ and $\tilde{\epsilon}(b_2)$ at somewhat higher b-values (where the b$^2$ terms are more dominant) and perform a separate, third acquisition for extracting MD at lower b-values. This 3-shell approach was thus preferred in this study.

# Materials and Methods

This study was carried out in accordance with the recommendations of the directive 2010/63/EU of the European Parliament of the Council, authorized by the Champalimaud Centre for the Unknown's Animal Welfare Body, and approved by the national competent authority (Direcção Geral de Alimentação e Veterinária, DGAV).

**Specimen Preparation.** Spinal cord specimens were obtained from adult male Wistar rats (N=2) weighing ~300 gr. The rats underwent standard transcardial perfusion under deep pentobarbital anesthesia. Cervical spinal cords were extracted, washed in PBS, and kept in 4% paraformaldehyde (PFA) for 24h at 4°C. The samples were then placed in freshly prepared phosphate buffer saline (PBS) for at least 48h prior to MRI experiments. The samples were cut to ~1cm lengths and placed in a 5 mm NMR tube filled with fluorinert (Sigma Aldrich, Lisbon, Pt).

**MRI experiments.** All MRI experiments were performed on a vertical 16.4T (700 MHz $^1$H frequency) Aeon Ascend scanner (Bruker, Karlsruhe, Germany) interfaced with a Bruker AVANCE IIIHD console. A Micro5 probe equipped with a 5 mm birdcage coil for transmit and receive functions and a gradient system capable of producing amplitudes of up to 3T/m isotropically was used. The sample was kept at a constant temperature of 23 °C throughout the experiments by means of air flow, and the samples were allowed to equilibrate with the surrounding temperature for at least 4 h before acquiring any diffusion or relaxation experiments.

All diffusion sequences were written in-house and were based on an Echo Planar Imaging (EPI) readout. For both DODE and DDE, the same acquisition parameters were used, namely, two-shot and double-sampled EPI with a readout bandwidth of 555.555 kHz, Field of View (FOV) of 6x4 mm$^2$ and in-plane matrix size of 120×80, leading to an isotropic in-plane resolution of 50×50 μm$^2$. The slice thickness was 500 μm, and TR/TE = 2500/52 ms. For both DODE and DDE acquisitions, Jespersen's 5-design sampling scheme (Jespersen et al., 2013) was used for the diffusion weighted images, and, additionally, eight images with zero b-value were acquired, such that the total number of images acquired in a given scan was 80. For both DODE and DDE, three separate acquisitions were performed with different b-values, namely, 2b = 1.2, 2.4 and 3.0 ms/μm$^2$ (where the factor of 2 reflects the accumulated diffusion weighting along the two diffusion epochs). The specific b-values were chosen based on signal-to-noise and contrast considerations: on the one hand, they have to be sufficiently low such that even higher-order terms do not contribute, but on the other hand, they have to be high enough for μFA contrast to be detectable. The lowest b-value scans were acquired with 12 averages, while the other two b-value shells were acquired with 32 averages each. The DODE diffusion parameters were: $T_{DODE}$ = 13 ms, N = 5, $\tau_s$ = 2 ms. The DDE diffusion parameters were Δ/δ = 12/1 ms, $\tau_m$ = 12 ms, see Figure 1 for definitions of the parameters.

Additional experiments were performed for mapping myelin water fraction. Those consisted of a Carr-Purcell-Meiboom-Gill (CPMG)-based acquisition performed using a modified pulse multi-slice-multi-echo (MSME) sequence. The same slice was acquired as in the diffusion images with identical in-plane resolution and FOV. The acquisition bandwidth for the pulse sequence was 100 kHz, and the pulses used for slice-selective excitation and refocusing had durations of 1.16 ms (Shinnar-Le-Roux shape) and 50 μs (Gaussian shape), respectively. The respective bandwidths of the excitation and refocusing pulses were 3625 Hz and 32100 Hz, respectively, such that the refocusing pulse provided complete refocusing on the entire slice. The ΔTE that could be achieved using these parameters was 2.85 ms, and 96 echoes were acquired from 2.85 to 273.6 ms. The repetition time was 2500 ms and two averages were acquired.

**Diffusion data preprocessing.** All preprocessing and analyses were performed using MatLab® (The MathWorks, Inc., Natick, Massachusetts, United States). Raw images were registered using an implementation of (Guizar-Sicairos et al., 2008) found in https://goo.gl/3bGU8b. The images were then denoised using Veraart's algorithm based on Marchenko-Pastur distributions in Principal Component Analysis of redundant data (Veraart et al., 2016b). Gibbs unringing was performed using Kellner's method (Kellner et al., 2016) implemented in Matlab . Finally, the denoised and unrung images were very slightly smoothed using a [2 2] median filter.

**Relaxation data preprocessing**. The preprocessing steps for the relaxation data were identical to the diffusion data preprocessing steps, except for an additional step in the very beginning of the pipeline whereby the magnitude data was converted to real data using Eichner's method (Eichner et al., 2015). All steps listed above including denoising, unringing and median filter smoothing were then executed in sequence.

**Diffusion data analysis.** The first analysis step for D(O)DE data was to fit the diffusion tensor. Diffusivities were computed using a simple linear fitting of $S_\parallel$ data acquired at the lowest b-value experiments followed by diagonalization and extraction of the diffusion tensor eignevalues. The mean diffusivity and fractional anisotropy were then calculated from the tensor eigenvalues as $MD = \frac{1}{3}(\lambda_1 + \lambda_2 + \lambda_3)$ and $FA = \sqrt{\frac{3}{2}\frac{(\lambda_1-MD)^2+(\lambda_2-MD)^2+(\lambda_3-MD)^2}{\lambda_1^2+\lambda_2^2+\lambda_3^2}}$, where $\lambda_i$ represent the tensor eigenvalues.

The second step in the analysis was to use the data from the two higher b-values to extract µFA. First, $\widetilde{\mu A^2}$ was extracted directly from Eq. 4; the mean diffusivity estimate was then used along with the extracted $\widetilde{\mu A^2}$ to obtain µFA via Eq. 2.

**Relaxation data analysis.** Following the preprocessing steps listed above, the filtered relaxation data were subject to a voxelwise inverse Laplace Transform (iLT) using 150 $T_2$ components log-spaced between 2.1 and 328.3 ms. The $T_2$ spectra were smoothed by minimum-curvature constraint as in (Dula et al., 2010) and extended phase graph analysis was performed to account for any $B_1^+$ inhomogeneity and ensuing stimulated echoes (Prasloski et al., 2012). The myelin water fraction (MWF) was computed from each spectrum as the fraction of signal originating from components with peak $T_2$ smaller than 17 ms. ROIs were drawn manually on the raw data closely following (Dula et al., 2010), and the ROI data was subjected to the same analysis using the mean signal decay in each ROI.

**Statistical analysis.** Gray matter and white matter masks were created by thresholding MWF maps with MWF<0.22 for gray matter and MWF>0.25 for white matter. The histograms in Figure 4 were then generated for each metric/method using Matlab's *histogram* function which automatically selects the bin width to represent the underlying distribution in the most accurate way. Parameter means and standard deviations are reported in the text and Tables.

Correlation analyses between different diffusion metrics were performed using automatic outlier rejection (Grubbs test for outliers) followed by calculation of Spearman's ρ (µFA and FA data from all methods were not normally distributed). An analysis of variance (ANOVA) was performed to compare µFA and FA arising from DODE and DDE methods, with post-hoc Bonferroni tests corrected for multiple comparisons.

To correlate MWF with µFA or FA extracted from the different methods, the diffusion maps were registered to the MWF using Matlab's *imregister* function using a *multimodal* configuration, initial radius of 1e-5, maximum number of iterations = 1000, and allowing for affine transformations due to the small differences in image geometry arising from EPI-based (diffusion) and line-by-line (relaxation) acquisitions.

When linear fits are presented (Figure 8), Matlab's *robustfit* function was used to extract the coefficients.

# Results

Diffusion data quality can be appraised in Figure 2, which plots representative raw data from one of the spinal cords, obtained from experiments with zero b-value (Fig. 2A), parallel (Fig. 2B), and perpendicular (Fig. 2C) diffusion orientations at the highest b-value used in this study. Before denoising, the worst-case signal to noise ratio (SNR) – measured at the highest b-value and with significant diffusion weighting gradients in the direction parallel to the spinal cord's principal axis – was ~20 in white matter. The middle column in Figure 2 shows the corresponding preprocessed data and the ensuing enhancement of image quality from denoising and Gibbs unringing (Figs 2D-F). Figures 2G-I show the result of subtracting raw and denoised images. The lack of structure in the subtracted images suggest that indeed only noise was removed and that no significant signal components were lost during denoising (Veraart et al., 2016b). The SNR of the preprocessed images was enhanced by a factor of ~2.

To assess the different maps obtained in this study, representative µFA and FA maps derived from DODE as well as DDE experiments (hereafter referred to as µFA$_{DODE}$ and µFA$_{DDE}$ or FA$_{DODE}$ and FA$_{DDE}$, respectively) are shown in Figure 3. Several interesting qualitative features can be highlighted from these images: (1) both µFA$_{DODE}$ and µFA$_{DDE}$ maps (Figs. 3A and 3C) have higher values than their FA$_{DODE}$ and FA$_{DDE}$ counterparts (Figs. 3B and 3D) in white matter, as well as in gray matter; (2) µFA$_{DDE}$ is higher and less tract-specific when compared with µFA$_{DODE}$ (for approximate definitions of tract locations and spinal cord anatomy, the reader is referred to Figure 3E); (3) µFA$_{DDE}$ appears quite homogeneous in the WM while µFA$_{DODE}$ shows more variation within WM; (4) similarly, FA$_{DDE}$ is more homogeneous in white matter compared with FA$_{DODE}$, which shows a greater variance in different tracts. To provide a more quantitative view on these features, Figure 4 plots histograms of µFA and FA in white matter and gray matter (c.f. Figure 4A and 4B for the ROI masks). In white matter, µFA$_{DODE}$ is higher than its FA$_{DODE}$ counterpart (Figure 4C), while in gray matter, µFA$_{DODE}$ is distributed at much higher values compared to FA$_{DODE}$ (Figure 4D). Similar trends were observed for DDE but with µFA or FA shifted towards somewhat higher values (Figures 4E and 4F).

It is also interesting to compare differences between methods within the same tissue type (e.g., comparing same-color distributions down the columns of Figure 4). µFA$_{DODE}$ is clearly lower and more widely distributed compared with µFA$_{DDE}$ in white matter. In gray matter, µFA$_{DDE}$ measured is high, while µFA$_{DODE}$ is somewhat smaller. Another interesting finding in gray matter, is that FA$_{DODE}$ and FA$_{DDE}$ values are only slightly different. The means and standard deviations of µFA and FA for each method are tabulated in Table 1.

A statistical analysis of these data is given in Figure 5, which presents box plots of the data. A one-way ANOVA revealed that in each tissue type (e.g., white matter or gray matter), all four metrics are highly statistically significantly different from each other (corrected p<1e-12, post-hoc Bonferroni test). However, it should be noted that although the metrics are different, they are not completely uncorrelated. Table 1 reports Spearman's ρ and its significance levels when comparing µFA and FA (extracted by the same method) in each ROI. While µFA$_{DODE}$ and FA$_{DODE}$ are correlated in white matter (Spearman's ρ = ~0.41), µFA$_{DDE}$ and FA$_{DDE}$ metrics are only weakly correlated (Spearman's ρ = ~0.19). In gray matter, the correlations between µFA and FA are weak for both methods and (Spearman's ρ = 0.22 and -0.10 for DODE and DDE, respectively). Note that although outlier rejection was used, in all cases less than ~1% of the data were identified as outliers and rejected.

To establish whether and how myelin modulates the anisotropy metrics, Carr-Purcell-Meiboom-Gill (CPMG) MRI experiments were performed on the same slice with the same resolution as the diffusion experiments. To assess the quality of the data, Figures 6A and 6B show the preprocessed data at short and very short TE of 2.9 ms and very long TE of 142.5 ms, respectively, in a representative spinal cord. Even at the very long TE, the SNR remains very high, especially after denoising. Denoising and unringing procedures were validated and found to have no negative impact on the quality of $T_2$ fitting procedure (data not shown), while improving the fits significantly. Figure 6C shows ROIs drawn in the major tracts of the spinal cord, while Figures 6D and 6E show the $T_2$ decays (with the ordinate drawn in log scale) and the resultant $T_2$ spectra (with the abscissa drawn in log scale), respectively. The decays in white matter are clearly non-linear, and the myelin water can be seen as an early peak in the $T_2$ spectrum with its peak $T_2$ around ~10 ms.

A representative myelin water fraction (MWF) map arising from pixel-by-pixel quantification of the spectra is shown in Figure 7A. Note the sharp contrast between the different tracts in MWF: for example, the dCST shows the lowest MWF (MWF ~ 0.30) while VST and FC exhibit the highest MWF (MWF ~ 0.45). Scatter plots between MWF and µFA or FA in white matter are shown in Figure 7 for DODE (Figs. 7B) and DDE (Figs. 7C), respectively. Table 2 summarizes the correlation coefficients and associated statistics. A moderate anticorrelation between MWF and µFA$_{DODE}$ is observed in the white matter (Spearman's $\rho = \sim -0.36$), while FA$_{DODE}$ did not correlate with MWF in a statistically significant manner. The DDE counterparts µFA$_{DDE}$ and FA$_{DDE}$ exhibited weak anti-correlation and correlation, respectively. Figures 7D and 7E show similar plots as described above, but for gray matter. Notably, correlations between MWF and FA$_{DODE}$, as well as FA$_{DDE}$ were very weak and their statistical significance not very high; on the contrary, µFA$_{DODE}$ was found to correlate somewhat with MWF, while µFA$_{DDE}$ correlated moderately with MWF, with very high statistical significance (c.f. Table 2).

Finally, the correlation of the mean µFA in the different tracts with literature regional averaged axon diameters was assessed. Figures 8A and 8B plot mean µFA and FA against the axon diameters reported in (Dula et al., 2010) for the different spinal cord tracts. These data, along with the values tabulated in Table 3, demonstrate that µFA$_{DODE}$ exhibits very strong anticorrelation with axon diameters (Spearman's $\rho = -0.96$, $p = 0.0028$). All other metrics are not significantly correlated with axon diameter.

## Discussion

µFA has been recently gaining increasing attention as a potentially useful source of contrast in microstructural MRI due to its ability to disentangle anisotropy from orientation dispersion. Methods other than D(O)DE, such as mapping µFA from experiments tailoring b-tensor shapes are emerging, with many potential applications (De Almeida Martins and Topgaard, 2016; Lasič et al., 2014; Szczepankiewicz et al., 2015; Westin et al., 2016). However, such methods may be confounded by time-dependent diffusion effects (De Swiet and Mitra, 1996; Jespersen et al., 2017c; Vellmer et al., 2017a, 2017b), whereas D(O)DE at long mixing times naturally avoid these confounds (Jespersen, 2012). It is therefore imperative to investigate how µFA may be correlated with underlying microstructural features such as axon dimensions and myelin, much like the early studies aiming to understand the sources for FA (Beaulieu, 2002; Kozlowski et al., 2008a; Mädler et al., 2008; West et al., 2016). In general, perhaps the most significant findings of prior studies on FA (conducted nearly invariably with SDE) were that (1) anisotropy in white matter depends on axonal membranes; and (2) the presence of myelin can further modulate FA metrics (Beaulieu, 2002). The application of oscillating gradients has also been shown to generate more contrast and more accurate estimations of small dimensions as compared to long diffusion time experiments, presumably due to the more efficient probing of smaller dimensions via the shorter diffusion times (Álvarez et al., 2013; Jiang et al., 2016; Xu et al., 2014).

The present study aimed to investigate how µFA differs from FA in terms of correlations with myelin water and axonal diameters, and to compare those metrics when measured with DDE or DODE sequences. We first focus attention to our results arising from white matter tissue. Notably, µFA was always larger than FA (Figures 3-5 and Table 2), in agreement with previous DDE experiments in fixed tissues (Jespersen et al., 2013) and in-vivo (Lawrenz et al., 2016). Since the µFA and FA metrics were extracted from the same acquisition, it is unlikely that other effects such as exchange or relaxation contributed to µFA>FA. Thus, our finding supports the notion that that orientation dispersion is significant even in highly structured tissues, such as spinal cord white matter. This is in excellent agreement with a recent study of SDE-derived diffusion tensor and kurtosis time-dependencies which also pointed to the same conclusion in pig spinal cord (Jespersen et al., 2017a), as well as with histological studies attempting to measure the dispersion directly in white matter (Leergaard et al., 2010). It is difficult to draw conclusions on whether the orientation dispersion arises within intra- or extra-axonal spaces (or both), or, whether undulations (Nilsson et al., 2012) or passing collateral fibers (Lundell et al., 2011) can contribute to these observations. Performing similar spectroscopic measurements utilizing cell-specific markers such as NAA or mI (Shemesh et al., 2014b, 2017), or performing much more extensive time/frequency/b-value-dependent measurements on water (Veraart et al., 2016a)(Papaioannou et al., 2017; Reynaud et al., 2016), or on metabolites (Palombo et al., 2016a; Valette et al., 2018) may further assist in addressing this question in the future.

Another interesting aspect when comparing µFA with FA in white matter, is that the two metrics are only moderately correlated when measured with DODE, and very weakly correlated when measured with DDE (c.f. Table 1). This finding suggests that when diffusion is encoded using oscillating gradients, spins experience less orientation dispersion than when they are probed using block gradients, since µFA would be perfectly correlated (and identical) to FA for perfectly aligned fibers. Hence, our findings point to specific length scales for orientation dispersion that are probed differently using the different sequences.

Next, we consider the relationships between myelin and µFA. Akin to its FA counterpart – µFA is ambiguous in that a compartment with length 'L' and radius 'R' can give rise to the same µFA as a compartment with length 2L and radius 2R. The axial path length could be restricted due to nodes of Ranvier, non-ideal cylindrical structure, varicosities, etc. However, if the path length parallel to the (assumingly) ellipsoids is constant, then one could predict that when larger amounts of myelin surround an axon, the µFA will be smaller as the restriction will increase in the perpendicular direction. However, in our study, a moderate *negative* correlation was observed between MWF and µFA$_{DODE}$ in white matter (Figure 7 and Table 2). This can be fully explained by considering the dependence of MWF and axon diameter via the g-ratio (Guy et al., 1989): the larger the axon, the thicker the myelin around it in (healthy) mammalian white matter (Innocenti, 2011). Hence, the negative correlation between µFA$_{DODE}$ and MWF would reflect indirectly the approximately constant g-ratio in healthy tissue, rather than enhanced restriction. Interestingly, µFA$_{DDE}$ showed a much weaker, yet still negative correlation with MWF. Since the microstructure has not changed between measurements, this likely reflects that DODE and DDE probe different path lengths *parallel* to the spinal cord's major axis: the larger the diffusion time, the longer path will be probed in the unrestricted dimension, and thence the µFA will be larger and less reflective of axon diameter or, by proxy, its myelin thickness. FA$_{DDE}$ showed a small positive correlation with MWF, which perhaps reflects the ambiguity of probing restriction and orientation distribution at the same time. Extracellular space contributions again cannot be neglected here, but for coherently aligned systems the arguments are similar as one could potentially treat the space between densely packed axons as potentially even more restricted than the intra-axonal space itself (Shemesh et al., 2011). It is also worth mentioning that MWF extracted from multiexponential T$_2$ measurements, as performed in this study, have been shown in the past to reflect microstructural metrics such as axon size and myelin thickness very faithfully in white matter (Dula et al., 2010; Kozlowski et al., 2008a; MacKay et al., 2006).

Our most striking findings in this study, perhaps, is that µFA$_{DODE}$ showed an extremely high, and statistically significant, negative correlation with axon diameters reported by (Harkins et al., 2016) for the different tracts (Table 3). This observation lends further credence to the explanation above: the finite parallel length scale probed by DODE makes the measurement strongly dependent on the perpendicular restriction, which in this case is reflected through axon sizes. Although the axon diameters were obtained from literature, it is worth stressing that axon diameter dependence in healthy spinal cords is highly reproducible and that the tracts analyzed were obtained from very similar cervical slices as in (Dula et al., 2010). Such a strong correlation is also highly unlikely to be obtained randomly. It is very interesting to also note that all other metrics did not correlate in a statistically significant fashion with axon diameters: µFA$_{DDE}$ likely due to its probing of longer parallel lengths, and the FA from both methods due to its inherent conflation or restriction with orientation dispersion.

In the spinal cord gray matter, very low FA$_{DODE}$ and FA$_{DDE}$ values were measured, suggesting a much lower degree of restriction compared to white matter diffusion. However, the µFA$_{DODE}$ and µFA$_{DDE}$ metrics in gray matter were still very high in the gray matter. In fact, the values reported in Table 1 also reflect $\frac{\mu FA_{DODE}^{GM}}{\mu FA_{DODE}^{WM}} = \sim 0.92$ and $\frac{\mu FA_{DDE}^{GM}}{\mu FA_{DDE}^{WM}} = \sim 0.89$. Combined with the low FA values in the gray matter, our findings suggest that a significant component of gray matter tissue experiences restricted diffusion but with a large degree of orientation dispersion. This finding is also in good agreement with previous literature demonstrating significant angular DDE modulations in ex-vivo gray matter (Shemesh et al., 2011; Shemesh and Cohen, 2011a). Further studies are needed to establish which underlying biological components give rise to such high µFA in gray matter, but dendrites, astrocyte branches, and nonmyelinated or myelinated axons traversing gray matter could be suspected (Palombo et al., 2016b, 2017). Time-dependent or spectroscopic experiments on metabolites could provide insight into such questions in the future.

Several limitations can be identified in this study. First, we have introduced a new way of measuring $\widetilde{\mu A}^2$ harnessing the 5-design acquisition at two b-values to reduce the recently-reported bias in $\mu A^2$ estimation due to higher order terms. Our new method is likely inferior to a sampling of a large range of b-values and the ensuing polynomial fitting as done in Ianus et al (Ianuş et al., 2017a). However, the advantage of the current approach is that it manages to avoid a prohibitively long experiment duration. Future studies will identify the accuracy and precision of the method proposed above vis-à-vis the ground-truth, and attempt to find optimal b-values for measuring $\widetilde{\mu A}^2$ as accurately and with as little bias as possible. Second, to compute µFA, we executed a third measurement at lower b-value to extract *MD*, which is then input into Eq. 2 along with $\widetilde{\mu A}^2$. However, MD itself may be conflated with higher-order terms, as pointed out recently by (Chuhutin et al., 2017); in this study, this issue was not accounted for, and may induce minor biases in the measurements of µFA. Better estimation of MD could probably be performed by sampling one or more low b-values and fitting kurtosis and MD at the same time from spherically averaged data. In addition, we have not explored the impact of specific b-value selection. At too low b-values, the difference in the log signals is very small, while at higher b-values, even higher-order terms may come into play. Third, the sample size was quite small (N = 2 spinal cords, only a single slice per cord), such that the variability across animals was not very well sampled. However, it is worth noting that the results were actually very consistent between both spinal cords: the mean µFA and FA, for both DODE and DDE, varied less than 10% between the cords (both in gray and white matter tissues), and the MWF varied less than 6% between the tissues. Although this consistency is promising for the robustness of the approach, the small number of samples renders this study perhaps more exploratory. Fourth, the experiments were performed at a relatively long TE of 52 ms. Given that the MWF was associated with $T_2$<20 ms and that the other water $T_2$s were distributed between ~20-60 ms, the experiments can be considered completely filtered for (directly contributing) myelin water, as $e^{\frac{-TE}{T_{2myelin}}} \sim 0.005$. Exchange between myelin water and intra/extra-axonal water is highly likely to occur, which may also confound the measurements, although it should be noted that at least for conventional DODE MRI, the relatively long TE is nearly unavoidable due to the necessity of non-negligible diffusion gradient waveform durations. Double-stimulated-echo approaches (Jerschow, A, Muller, 1997; Shemesh and Cohen, 2011b) would thus be nearly impossible to execute for DODE, even before considering the significant SNR reduction associated with such sequences, $(1/2)^N$, where N is the number of stimulated echoes. Finally, a histological study was not here performed, and the study relies on literature reports of correlations between MRI-derived MWF and myelin thickness and the values for axon diameters. Future studies can expand the findings here and perform more direct correlations with histology, although it should be pointed out that big differences in these parameters are unlikely to be observed for healthy tissues. In addition, it would be fruitful to modulate the microstructure actively and to observe how µFA varies, e.g., using genetic mutations that alter myelin content. All these highly interesting avenues will be pursued in the future, but the present study provides the first steps in this direction.

# Conclusions

This study investigated the microstructural correlates of μFA and FA using high resolution D(O)DE experiments in fixed spinal cords at 16.4 T. Our results indicate very strong anticorrelations of μFA$_{DODE}$ with axon size, and moderate anticorrelations of μFA$_{DODE}$ with MWF, whereas μFA$_{DDE}$, FA$_{DODE}$ and FA$_{DDE}$ correlate to a much lesser or no extent with those microstructural features. These findings shed light on the mechanisms of restriction in spinal cord white matter when investigate without conflation by orientation dispersion. The correlations of μFA$_{DODE}$ with axon diameters and myelin water fraction are thus promising for future investigations of longitudinal variations in these properties, e.g., in disease or with learning.

## Acknowledgements

This study was funded in part by the European Research Council (ERC) under the European Union's Horizon 2020 research and innovation programme (grant agreement No. 679058 - DIRECT-fMRI). The author would like to thank Prof. Sune N Jespersen and Mr. Jonas Lynge Olesen (Aarhus University) for providing the codes for denoising and Gibbs unringing, as well as for many insightful discussions. The author also thanks Dr. Daniel Nunes for extracting the tissues used in this study, Dr. Andrada Ianus and Ms. Teresa Serradas Duarte for providing parts of code used in the analyses performed here, and Prof. Mark D Does from Vanderbilt University for the REMMI pulse sequence and its analysis tools, that were supported through grant number NIH EB019980.

## Author contribution statement

NS designed the study, collected and analyzed data, and wrote the paper.

## Conflicts of Interest

The author declares no conflict of interest.

# Figures

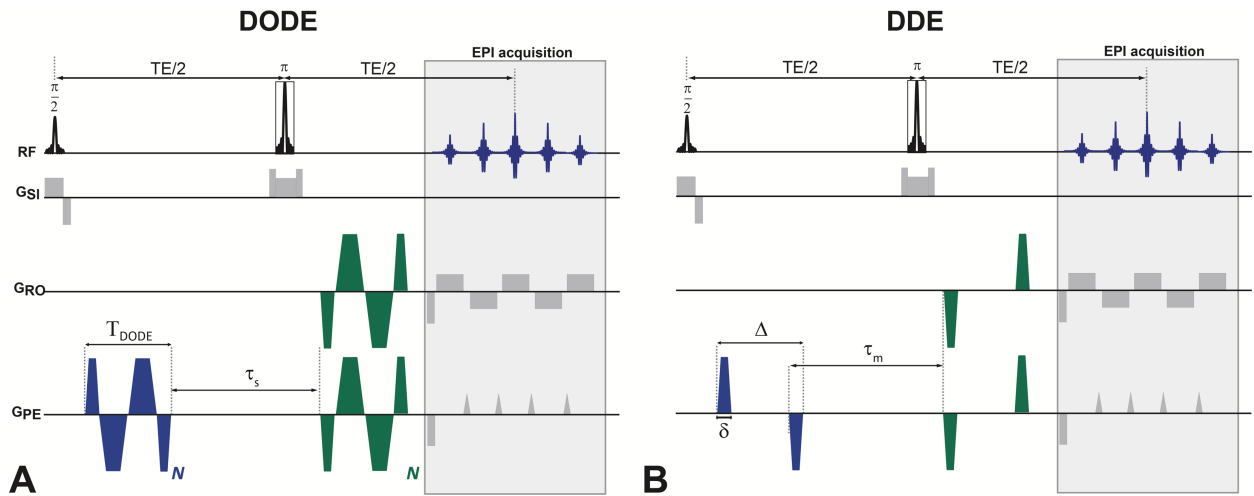

**Figure 1.** Diffusion MRI pulse sequences used in this study. **(A)** DODE and **(B)** DDE weightings were overlaid on a basic SE-EPI sequence. The diffusion gradient orientations are independent and can vary in any of the axes, the particular instantiation here represents one particular case where **G$_1$** is oriented along the PE axis and **G$_2$** is at an angle in the PE-RO plane. Other than the relative orientations that varied, identical waveforms were used for the two diffusion encodings.

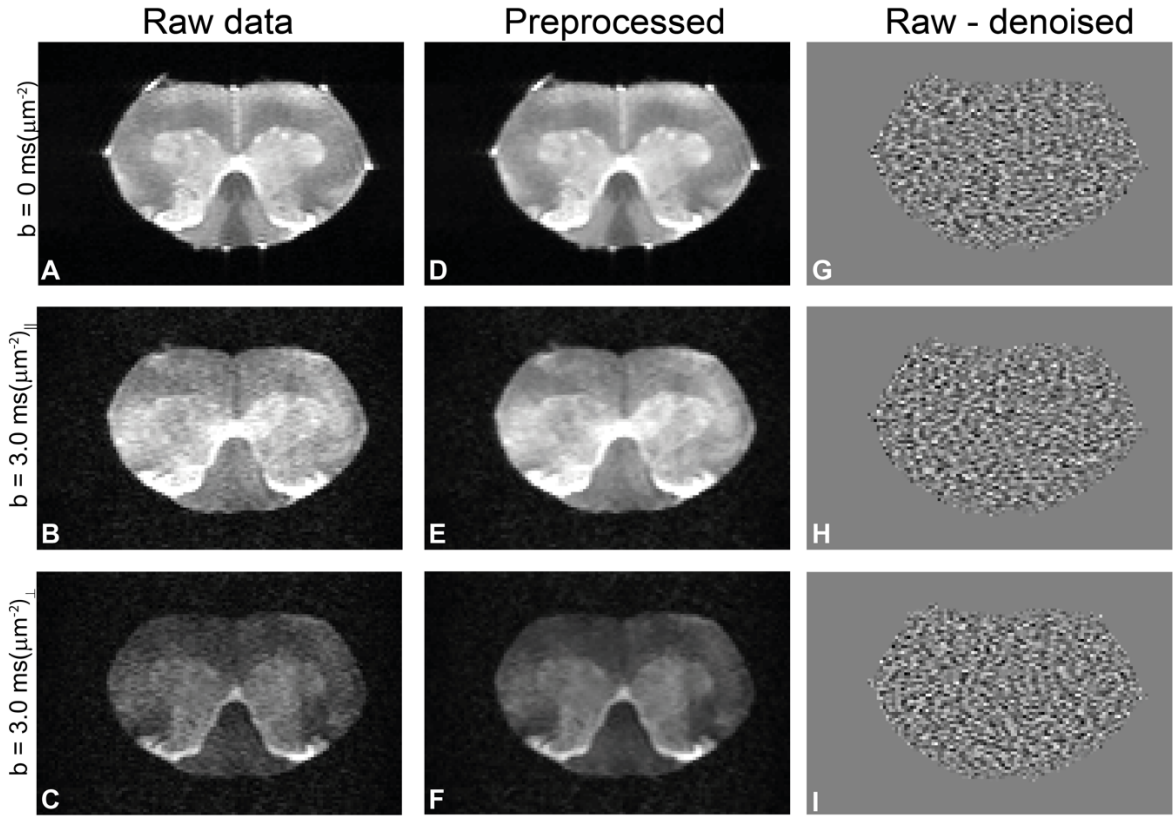

**Figure 2.** Quality of diffusion MRI data and preprocessing in a representative spinal cord. **(A-C)** Raw data with zero b-value, parallel, and perpendicular waveforms acquired at the highest b-value, respectively. The perpendicular waveform had more significant components along the spinal cord principal axis and thus show greater attenuation. **(D-F)** Results of preprocessing the data in A-C (denoising and Gibbs unringing). Notice how the noise is highly reduced in the preprocessed images without adverse effects to image quality. **(G-I)** Subtraction of denoised and raw data, showing only noise and thus demonstrating that no significant signal components were removed during Marchenko-Pastur PCA denoising.

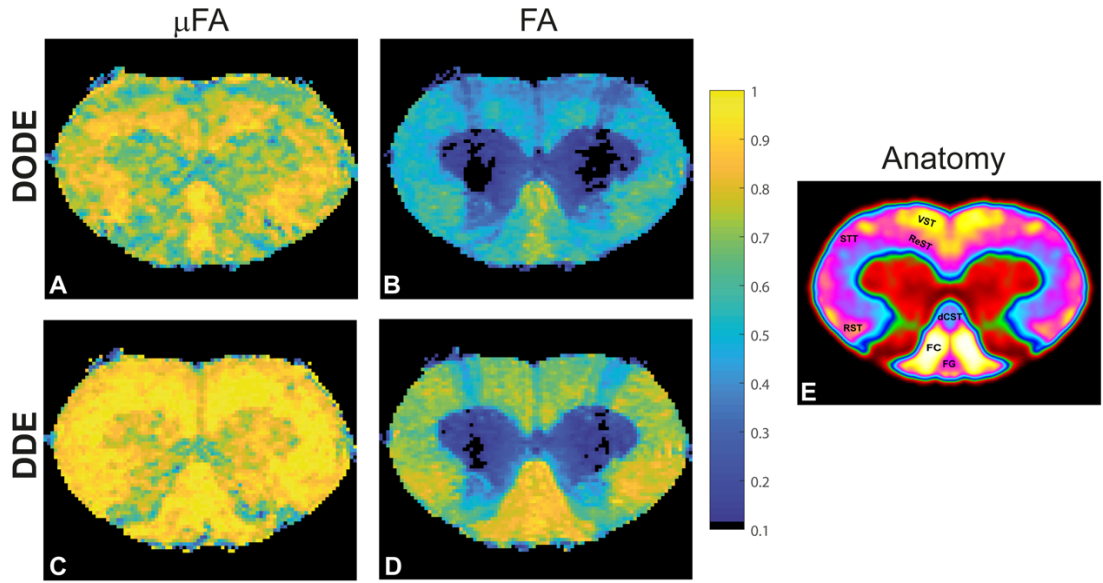

**Figure 3.** Parameter maps for a representative spinal cord. **(A)** µFA$_{DODE}$; **(B)** FA$_{DODE}$; **(C)** µFA$_{DDE}$; **(D)** FA$_{DDE}$. Notice the differences in contrast both in white and in gray matter tissues both between metrics and between sequences. Most notably, µFA is higher than FA and DDE-driven metrics are higher than DODE-driven metrics, especially in white matter. **(E)** Anatomy of the spinal cord for reference, displayed over a smoothed false-color image of the cervical segment. The gray matter is shown in red and green, while the tracts are highlighted on the left side of the cord.

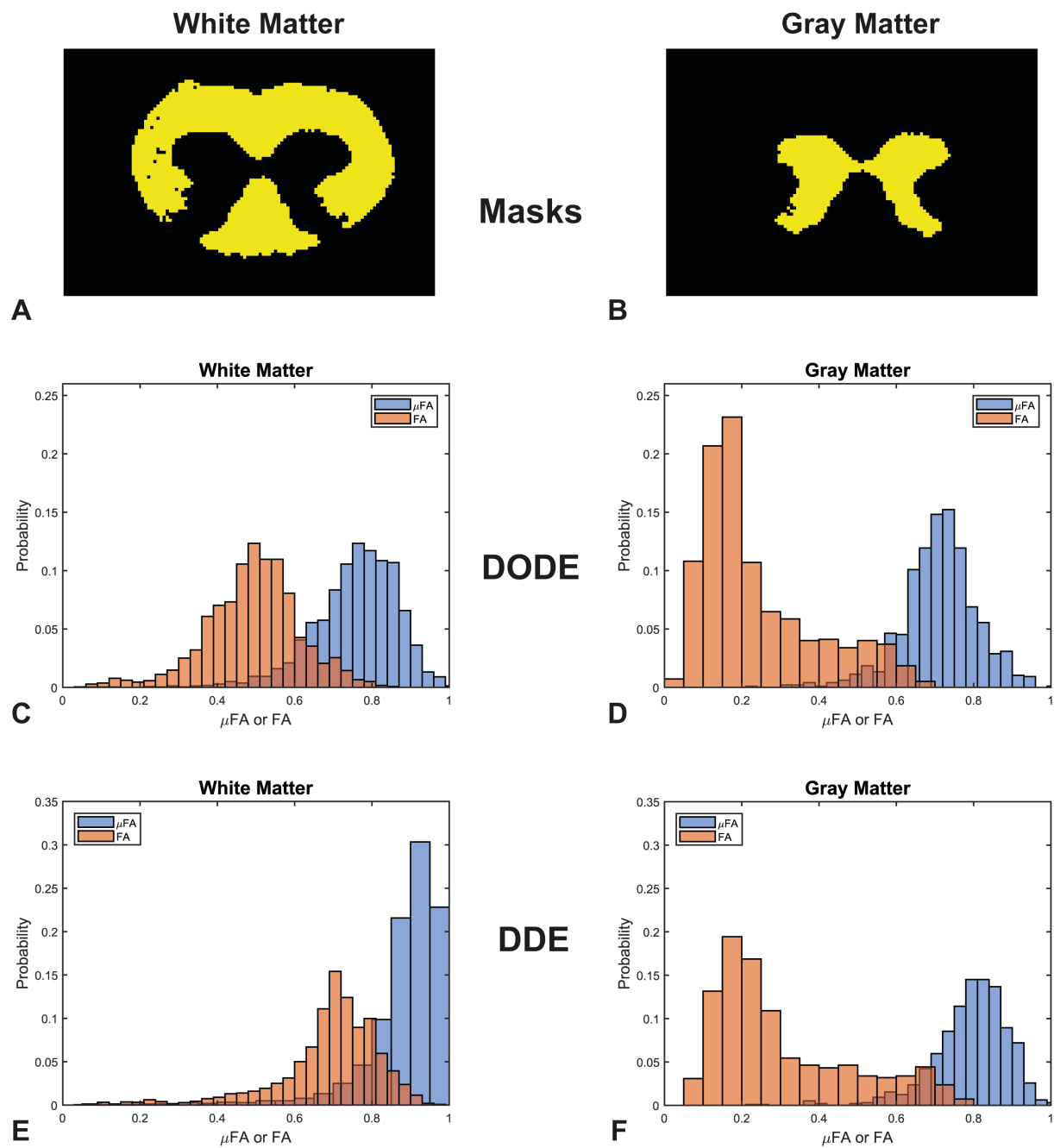

**Figure 4.** Histogram distributions of the different metrics in white matter and gray matter. **(A-B)** Masks for the white and gray matter tissues, respectively. **(C-D)** µFA$_{DODE}$ and FA$_{DODE}$ for white and gray matter. **(E-F)** µFA$_{DDE}$ and FA$_{DDE}$ for white and gray matter. Notice the different distributions in white matter for both DODE- and DDE-driven metrics, as well as the higher µFA as compared to FA in all tissues.

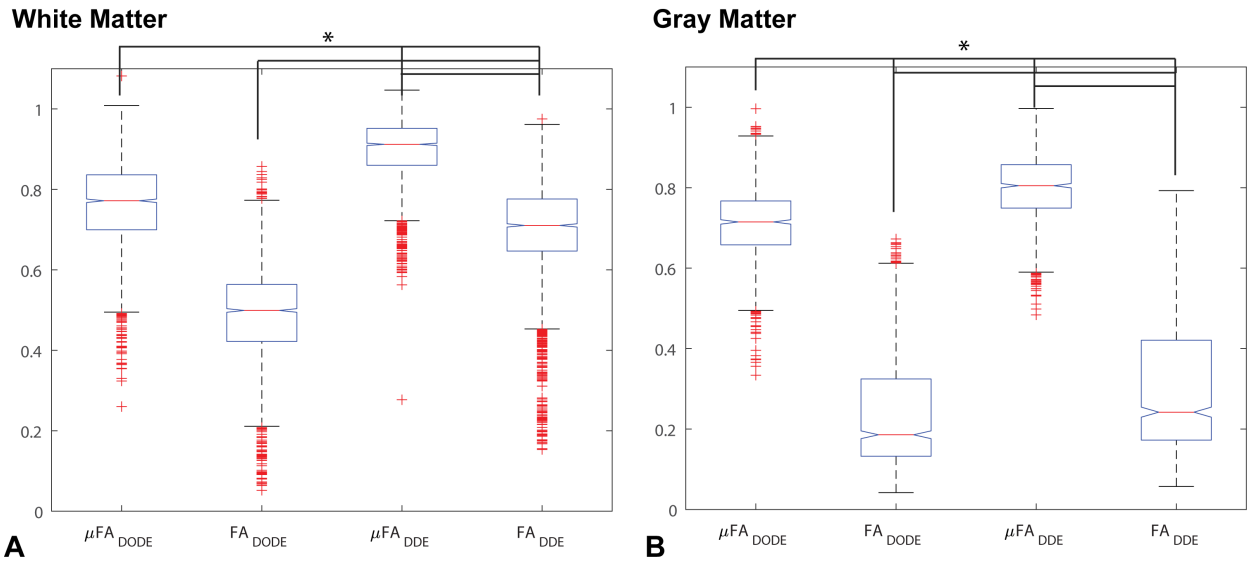

**Figure 5.** Box-and-whisker plots of the different metrics. **(A)** White matter analysis. **(B)** Gray matter analysis. *$p<10^{-12}$ between all pairs from ANOVA with Bonferroni post-hoc comparison and corrected for multiple comparisons.

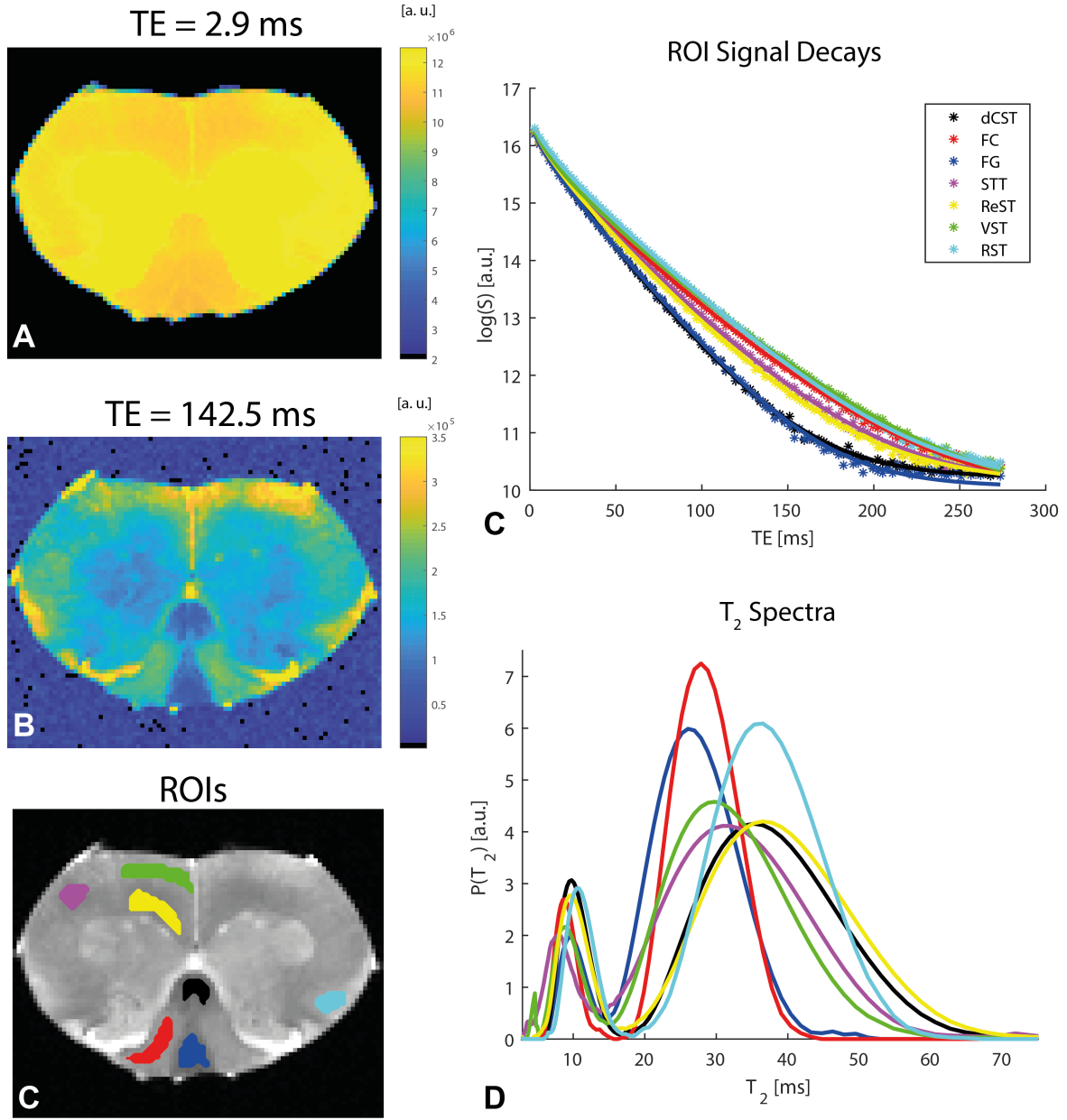

**Figure 6.** Relaxation data and analysis in a representative spinal cord. **(A-B)** Preprocessed data at short and long TEs, respectively, reveal excellent SNR. **(C)** ROI definitions. **(D)** Mean ROI signal decays with TE (symbols) along with their respective fits (solid lines). N.b. the log scale in the ordinate. **(E)** $T_2$ spectra (plotted in log scale in the abscissa) extracted from an iLT fit to the ROI data. The myelin water is associated with the peak corresponding to shorter $T_2$ values. The ROI colors in (C) correspond to the color of the plots in (D-E).

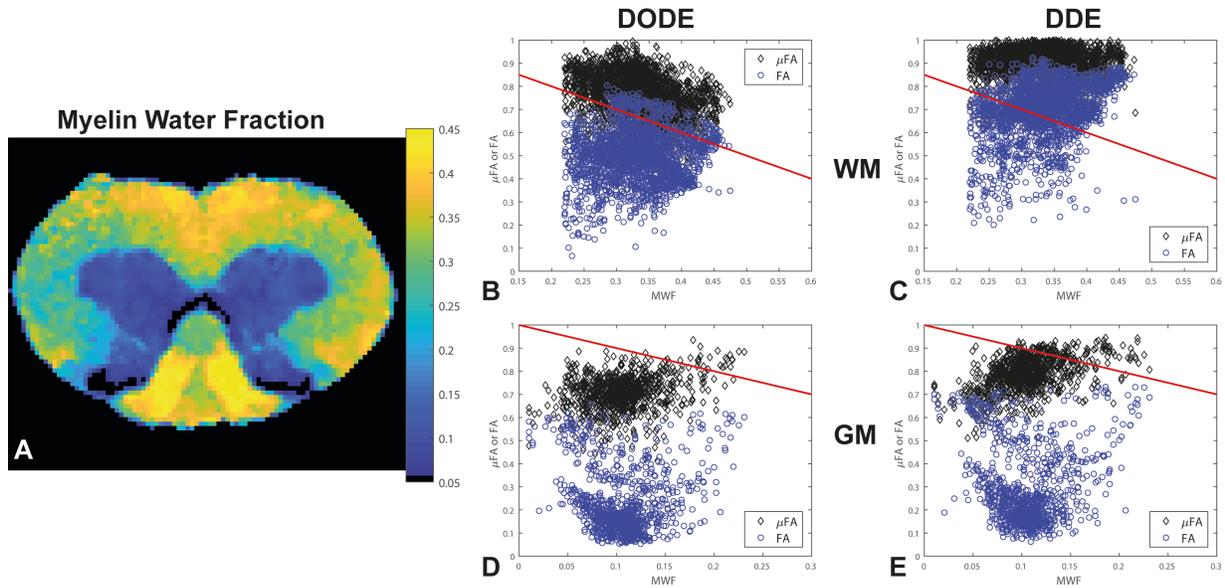

**Figure 7.** Myelin Water Fraction (MWF) and its correlations with diffusion-derived metrics. **(A)** MWF from a representative spinal cord, showing excellent contrast between the white matter and gray matter as well as within most white matter tracts. **(B-E)** Correlations between DODE and DDE metrics with MWF in white and gray matter tissues. Blue circles represent FA whereas black diamonds represent µFA. Red lines represent -1*identity to guide the eye.

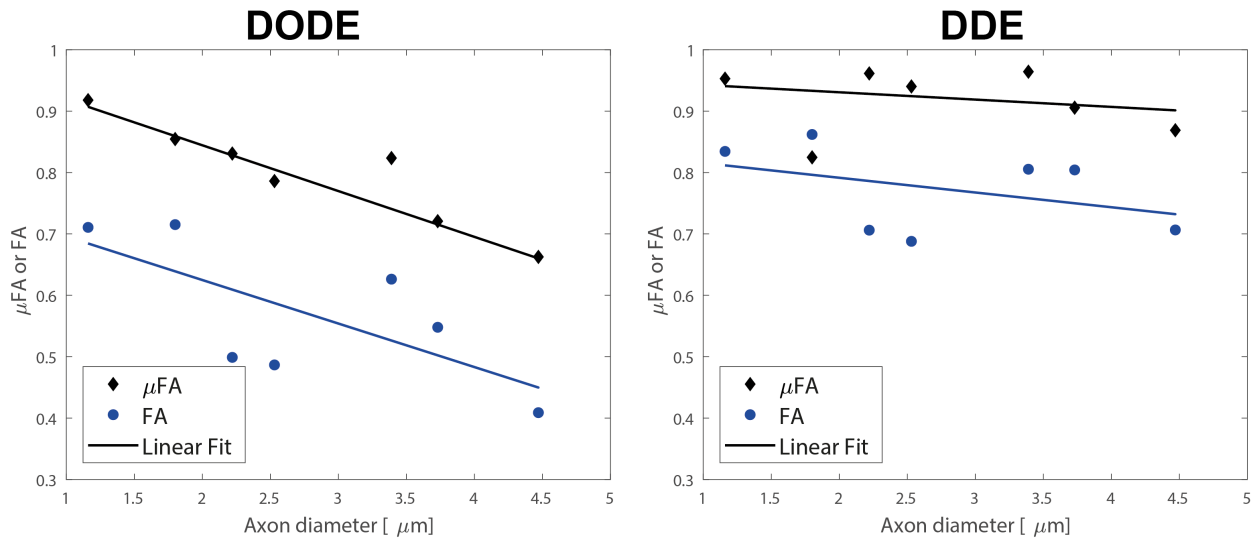

**Figure 8.** Correlations of diffusion metrics with literature-based average axon diameters in the different white matter tracts. **(A)** Correlations of metrics derived from DODE. **(B)** Correlations derived from DDE. The best linear regressions to the experimental data are also given as solid lines. Note the excellent inverse agreement between µFA$_{DODE}$ and axon diameter, which also had a very high anticorrelation coefficient of $\rho \sim -0.96$ while all other metrics did not show significant correlations.

# Tables

**Table 1.** White Matter and Gray Matter microscopic and fractional anisotropies, along with their Spearman correlation coefficient and significance.

|  | µFA DODE | FA DODE | Spearman's ρ | p-value | µFA DDE | FA DDE | Spearman's ρ | p-value |
|---|---|---|---|---|---|---|---|---|
| **White Matter** |  |  | 0.41 | $<10^{-10}$ |  |  | 0.19 | $<10^{-10}$ |
| Mean | 0.77 | 0.49 |  |  | 0.89 | 0.69 |  |  |
| σ | 0.10 | 0.12 |  |  | 0.16 | 0.13 |  |  |
| **Gray Matter** |  |  | 0.22 | $<10^{-10}$ |  |  | -0.10 | $<0.002$ |
| Mean | 0.71 | 0.24 |  |  | 0.79 | 0.31 |  |  |
| σ | 0.10 | 0.15 |  |  | 0.10 | 0.18 |  |  |

**Table 2.** Statistical analysis of correlations between (µ)FA and myelin water fraction in white matter and gray matter.

|  | µFA DODE | FA DODE | µFA DDE | FA DDE |
|---|---|---|---|---|
| **White Matter** |  |  |  |  |
| Spearman's ρ | -0.36 | 0.02 | -0.07 | 0.30 |
| p-value | $<10^{-10}$ | NS | 0.0011 | $<10^{-10}$ |
| **Gray Matter** |  |  |  |  |
| Spearman's ρ | 0.23 | 0.11 | 0.45 | -0.1 |
| p-value | $<10^{-10}$ | 0.0002 | $<10^{-10}$ | 0.0015 |

**Table 3.** Statistical analysis of correlations between (µ)FA and literature-based average axon diameter (extracted from Dula et al) in the rat spinal cord.

|  | µFA DODE | FA DODE | µFA DDE | FA DDE |
|---|---|---|---|---|
| **White Matter** |  |  |  |  |
| Spearman's ρ | -0.96 | -0.68 | -0.14 | -0.43 |
| p-value | 0.0028 | NS | NS | NS |